\documentclass[twocolumn,showpacs,preprintnumbers,amsmath,amssymb]{revtex4}
\usepackage{dcolumn} 
\usepackage{bm}      
\usepackage{hyperref}
\usepackage{longtable}
\usepackage{graphicx} 

\begin{document}

\title[Thermoelectric properties of molecular nanostructures]
{Thermoelectric properties of molecular nanostructures}

\author{Vladimir N. Ermakov}
\author{Sergei P. Kruchinin}
\email{skruchin@i.com.ua}
\affiliation{Bogolyubov Institute for Theoretical Physics, NASU,
Kiev 03680, Ukraine}

\author{Hyun Taki  Kim}
\affiliation{Metal-Insulator  Transition  Lab.,
Electronics and Telecommunications Research Institute,
Daejeon, 305-350,
Republic of Korea}

\author{Thomas Pruschke}

\affiliation{Institut f\"ur Theoretische Physik, Georg-August-Universit\"at
G\"ottingen, Friedrich-Hund-Platz 1, 37077 G\"ottingen, Germany}

\begin{abstract}
We use the concept of resonant tunneling to calculate the
thermopower of molecular nanosystems.  It turns out that the sign of the 
thermovoltage under resonant tunneling conditions depends sensitively on the participating molecular orbital, and one finds a sign change when the transport channel switches from the highest occupied molecular orbital to the lowest unoccupied molecular orbital.  Comparing our results to recent experimental data obtained for a BDT molecule contacted with an STM tip, we observe good agreement.
\end{abstract}

\maketitle

\section{Introduction}

Studies of the origin of a voltage or current in nanosystems in the presence of a temperature gradient are an extremely interesting
and promising area in the field of nanotechnologies
\cite{1,2,3}. 
There are several important possible future applications in several areas of devices, among them
 the development of nanothermosensors (see for example \cite{10}), which is especially
urgent for a number of technological processes and for
research in biology concerning the functioning of life.

However, different from the classical description
of thermoelectric phenomena, which is already challenging enough, the  necessity to apply strictly
quantum-mechanical methods in the realm of nano-objects makes the whole problem an extremely difficult one, and a proper theory for studying transport phenomena in the most general setup does not yet exist. 
However, for the description of most experimental realizations of thermoelectric transport through nanon-structures, one can fortunately make some simplifying assumptions. Usually, one can consider the system to consist of two metallic structures, which are typically very good conductors and which we will call leads, that are spatially separated. Hence, there will be no current flowing between the leads. Placing an active element like a molecule between these leads will thus induce a transport path and, when voltage or temperature differences between the leads are imposed, to thermoelectric phenomena \cite{overview}. The coupling of the molecule and the leads will  be of tunneling type, i.e.\ one can usually assume that this coupling is rather weak.

Although this setup seems rather straightforward, its actual experimental realization is by no means trivial. In particular, a good control over if a molecule is attached to both leads at all and how good the coupling to the leads is, is very hard to obtain \cite{2}. Furthermore, when using two extended leads on a common substrate, the introduction of temperature gradients poses a further challenge. A certain breakthrough along these lines has been achieved recently by using an STM tip to controlled pick molecules on a metal surface and generate well-defined break junctions. Here, certain control over the coupling between the molecule and the leads can be performed by moving the tip, and a temperature gradient can easily be applied by heating the metal while keeping the tip a fixed temperature \cite{2}. 


In the present paper, we develop a non-equilibrium description of the stationary thermotransport through such a nanostructure, using the idea of resonant tunneling. The model is rather simple, but can be solved analytically, and the theoretical results can be directly compared with experimental data. We find a rather good agreement, which we believe is in favor of this simple and straightforward model.

\section{Resonant tunneling with applied temperature gradient}

\begin{figure*}
\includegraphics[width=\textwidth]{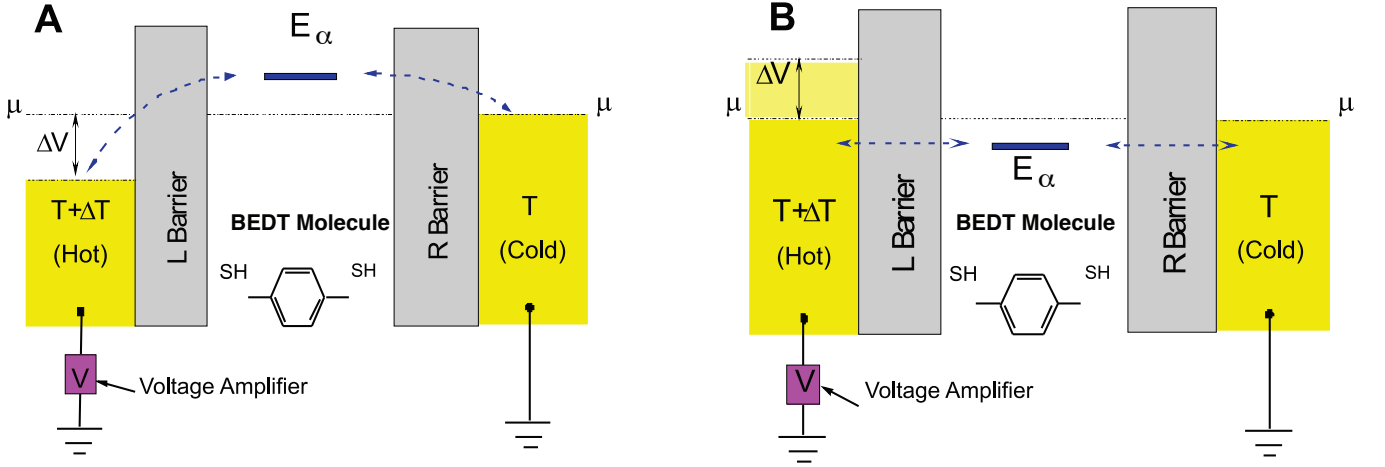}
\caption{Schematic model to study thermoelectric phenomena for a molecule attached to leads. 
{\bf A} shows the case where the LUMO provides the resonant level, and {\bf B} corresponds to a HOMO resonant level.\label{fig:1} The energies of orbitals and bias $\Delta V$ are chosen to provide a setup for thermopower measurment, i.e.\ no electrical current flowing.}
\end{figure*}

Resonant electron tunneling of particles through a system of double
potential barriers is very sensitive to a position of electronic
states in a quantum well \cite{4}. This circumstance can be used for
the effective control of the tunneling process. 
As a model of a double-barrier tunneling system, we assume the structure with the energy profile shown
schematically in Fig.\  \ref{fig:1} \cite{5,6,7,8,9}. The Hamiltonian describing the tunneling of
electrons through such a structure can be chosen in the form
\begin{equation}
\label{eq1} { H}={ H}_0 +{ H}_W +{ H}_T .
\end{equation}

\noindent The first term of this Hamiltonian is
\begin{equation}
H_0 =\sum\limits_{k\sigma } {\varepsilon _L (k)a_{k,L,\sigma }^\dagger } a_{k,L,\sigma
}^{\phantom{\dagger}} +\sum_{k\sigma } {\varepsilon _R (k)a_{k,R,\sigma }^\dagger} a_{k,R,\sigma}^{\phantom{\dagger}}
.
\end{equation}
It describes electrons in the left electrode (source) and in the
right  (drain). Because we are not interested in the detailed properties of the leads, we assume that these charge carriers can be taken to be quasi-particles, and $a_{k,L,\sigma }^\dagger (a_{k,L,\sigma }^{\phantom{\dagger}} )$ respectively
$a_{k,R,\sigma }^\dagger (a_{k,R,\sigma }^{\phantom{\dagger}} )$ are the creation (annihilation)
operators for these quasi-particle in  source respectively drain. The dispersions are in the same spirit given by
$\varepsilon _{L/R} (k)=\hbar ^2k^2/2m_{L/R}$, where
$m_{L/R}$ denote the effective masses for the left and right lead. We will assume in the following $m_L=m_R=m$ for simplicity.

The Hamiltonian $H_{W}$  describes the electronic states in the nano-object. It
can be written in the form
\begin{equation}
H_W =\sum_\alpha E_\alpha \,a_\alpha ^\dagger a_\alpha^{\phantom{\dagger}}+H_\text{I}\;,
\end{equation}
where $\alpha $ labels the single-particle levels of the molecule, and $H_\text{I}$ denotes possible interactions.
The energy in the well depends on the applied bias $\Delta V$ and can be written as $E_\alpha =\epsilon
_\alpha -e_0\,\gamma\, \Delta V$, where $\epsilon _{\alpha}$ is the bare energy of the
resonant state in the quantum well, $\Delta V$ the potential drop across the molecule, $e_0>0$ the elementary charge and $\gamma$ a factor
depending on the profile of the potential barriers (for identical
barriers, $\gamma  = 0.5$). Finally, the Hamiltonian $H_{T}$ describing the
tunneling of electrons through the barriers has the conventional
form
\begin{equation}
H_T =\sum_{k\alpha,\delta=L,R } \left({T_{\alpha\sigma}(k)\, a_{k,\delta,\sigma }^\dagger a_\alpha^{\phantom{\dagger}} } +\text{h.c.}\right)
\end{equation}
Here, $T_{\alpha\sigma}(k)$ is the matrix element of tunneling from
source respectively drain to and from the molecule. As usual, we assume that both are the same and that they also do not depend on the applied bias.

When we apply a constant external bias across the system, a
nonequilibrium steady-state electron distribution will result. We assume
that the electron distribution functions in the electrodes (source,
drain) are equilibrium ones, i.e.\ Fermi functions, due to the large volumes of these reservoirs, but their
chemical potentials and temperatures can be different. The chemical
potentials usually encode a voltage drop across the nano-region, hence in our model $\mu _L =\mu +\Delta\mu$, $\mu_R=\mu$ and $\Delta\mu=e_0\Delta V$.

The simple setup of the system in Fig.\  \ref{fig:1} makes the evaluation of nonequilibrium properties comparatively simple.The important quantity entering all formula is the density of states (DOS) $\rho(E)$ for the local level in the presence of the leads \cite{Meir_Wingreen}. To calculate it, we need the retarded Green's function $G_{\alpha,\alpha}(E)$ \cite{Mahan}, from which we can obtain the DOS as
%
\[
\rho (E)=-\frac{1}{\pi }\sum_\alpha {{\rm Im}\,G_{\alpha ,\alpha}(E)} ,
\]

The electron distribution function $g(E)$ in the quantum well is essentially
nonequilibrium. It can be determined from the condition of equality of the
tunneling currents through the source and the drain. Then the distribution
function has the form
\begin{eqnarray}
g(E)&=&\frac{1}{\Gamma (E)}[\Gamma _L (E)f_L (E)+\Gamma _R (E)f_R
(E)]\\
\Gamma (E)&=&\Gamma _L (E)+\Gamma _R (E)\;,
\end{eqnarray}
where $\Gamma _{L}$ and $\Gamma _{R }$ are the tunneling rates for source (L) and drain (R), given by the expressions \cite{Meir_Wingreen}
\begin{eqnarray}
 \Gamma _L(E) &=& \sum\limits_{k\sigma} {\vert T_{\alpha\sigma}(k) \vert } ^2\delta
(E-\varepsilon _L(k))\\
\Gamma _R(E)  &=& \sum\limits_{k\sigma} {\vert T_{\alpha\sigma}(k) \vert } ^2\delta
(E-\varepsilon _R (k))\; ,
\end{eqnarray}
and $ f_{L}(E)$ and $f_{R }(E)$ are the quasi-particle distribution functions in the
source and the drain, respectively. They have Fermi-Dirac form and read
temperature difference
\[
f_{L/R}(E) =\left\{ {\exp \frac{E-\mu _{L/R} }{k_B T_{L/R} }+1} \right\}^{-1},
\]
where $k_{B}$ is the Boltzmann constant, and $T_{L/R}$ are temperatures in the
source and the drain, respectively.

\section{Double barriers thermostructures for resonant tunneling}
With the above formula for the distribution function, one can straightforwardly evaluate physical quantitites.
For example, the occupancy of the molecule can be determined with the help
of expression \cite{5}
\[
n_\alpha =-\frac{1}{\pi }\int {dE\,g(E)\,{\rm Im}G(\alpha ,\alpha
,E)} .
\]
Moreover, the net current $J_{sd}$ between source and drain through the molecule is given by the equation \cite{7,Meir_Wingreen}
\begin{equation}
\label{eq2} J_{sd} =\frac{e_0}{\hbar }\int {G(E)\,(f_L -f_R )} \,\rho
(E)\,dE ,
\end{equation}
where $G(E)=\Gamma _L (E)\Gamma
_R (E)/\Gamma (E)$. Since we assume that the contacts are of tunnel type, the transition rates
$\Gamma _{R}$, $\Gamma _{L }$ are exponentially
dependent on the barrier widths and heights.
Correlation effects between the electrons in a nano-structure encoded in $H_\text{I}$ can be taken into account
by means of $\rho(E)$, too \cite{Meir_Wingreen}, and will in general dramatically modify the properties \cite{goldhaber-gordon,cronenwett}. The theoretical description of this situation is at present possible in the linear response regime only (see Mravile and Ramsak \cite{ramsak} for an overview). 

At this point one can only proceed analytically by assuming a
low transparency of the barriers, i.e.\ $\Gamma(E)\to0$, and a simple energy level structure, for example a quasi-particle description of the electronic structure based on first-principle methods. The
density of states $\rho(E)$ then only depends on the energy structure of the nano-structure via
\[
\rho (E)=\sum_\alpha C_\alpha \,\delta(E-E_\alpha),
\]
where $\delta(x)$ is Dirac's function and $C_{\alpha}$ a
weighting factor for the energy level $E_{\alpha}$.
For the current we then obtain the formula
\begin{equation}
\label{eq4} J_{sd}=\frac{e_0}{\hbar}\sum_\alpha C_\alpha \,G(E_\alpha )\,(f_L (E_\alpha )-f_R (E_\alpha ))\,. 
\end{equation}
The distribution functions $f_{L/R }(E_{\alpha})$ are exponentially
dependent on the energy $E_{\alpha}$. Thus, when $\vert E_\alpha -E_{\alpha'}
\vert \gg k_B\, T$, where $E_{\alpha'}$ denotes a neighboring molecular orbital, there will be one particular energy $E_{\alpha}$ for which $\vert E_\alpha -\mu \vert $ is minimal. With respect to this orbital, transport through all other will be exponentially suppressed, hence we need to keep only this particular orbital in the calculations. In this approximation, \ref{eq4} can be reduced to the form $G(E_\alpha )[f_L (E_\alpha )-f_R (E_\alpha)]=0$, respectively, for small but finite $\Gamma_{L/R}(E)$,
\begin{equation}
\label{eq5}
f_L (E_\alpha)-f_R (E_\alpha )=0\,.
\end{equation}
According to our definition, $E_{\alpha}$ is one of the energy levels of a BDT molecule which has the
smallest distance from the chemical potential, which will either be the highest occupied molecular orbital (HOMO) or the lowest unoccupied molecular orbital (LUMO). That molecule level is, as noted before,
shifted by the voltage $\Delta V$ by a value $\Delta E =-e_0\gamma\Delta V=-\gamma
\Delta\mu$. For an asymmetric barriere we have $\gamma=\Gamma_R(E_\alpha)/\Gamma(E_\alpha)$. Note that since the tunneling matrix elements for source and drain were chosen to be the same, this expression remains well-defined in the limit $\Gamma_{L/R}(E)\to0$.
The solution of \ref{eq5} then becomes
\begin{equation}
\label{eq6} e_0\Delta V=\frac{(E_{\alpha} -\mu )}{1+\frac{\Delta
T}{T}\frac{\Gamma_R }{\Gamma}}\frac{\Delta T}{T} .
\end{equation}
where as final step we also assumed that $\Gamma_{L/R}(E)\approx\Gamma_{L/R}$ for energies close to $\mu$.

\begin{figure}[t]
\includegraphics[width=6.5cm]{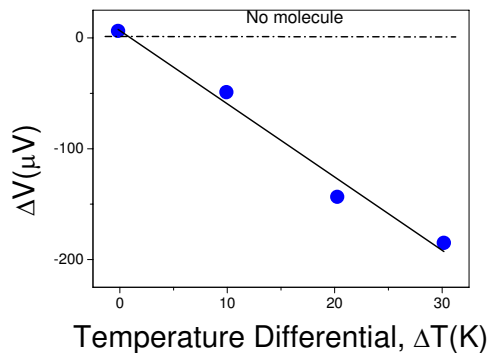}
\caption{Comparison of experimental data (blue circles \cite{2})
with theoretical curve (by \ref{eq6}) is demonstrated.
}\label{fig:2}
\end{figure}

\noindent Our  \ref{eq6} can now be compared with experimental data. In
particular, such a comparison with the experiments on BDT with STM \cite{2} results in
$\Delta V<0$. Acording to our setup in Fig.\ \ref{fig:1} this means that $E_{\alpha} <\mu $. Thus, the energy level
$E_{\alpha}$ seen in experiment is a HOMO, i.e.
the condition shown in  Fig.\ \ref{fig:1}B. In
fact, the conductivity of the tunneling structure is determined by
holes. The detailed comparison of the thermo-voltage dependence on
$\Delta V$ is demonstrated in Fig.\  \ref{fig:2}. In the case of $\Delta T
\ll T$, \ref{eq6} can be approximated by a linear relation
\begin{equation}
\label{eq7} e_0\Delta V=(E_\alpha -\mu )\Delta T/T .
\end{equation}
From the comparison with the experiment, we get $\mu -E_\alpha \approx
2$meV. 
%

To study the dependence of  thermoelectric effects on the distance $d$ between
substrate surface (source) and STM tip (drain), let us  consider our double-barrier system as a simple resistor network. In this case,  \ref{eq5} and \ref{eq4} can be
reduced to
\begin{equation}
\label{eq8} \frac{e_0}{\hbar }\,C_\alpha \,G(E_\alpha )[f_L (E_\alpha )-f_R (E_\alpha
)]=J_{sd} ,
\end{equation}
\begin{figure}[t]
\includegraphics[width=7.5cm]{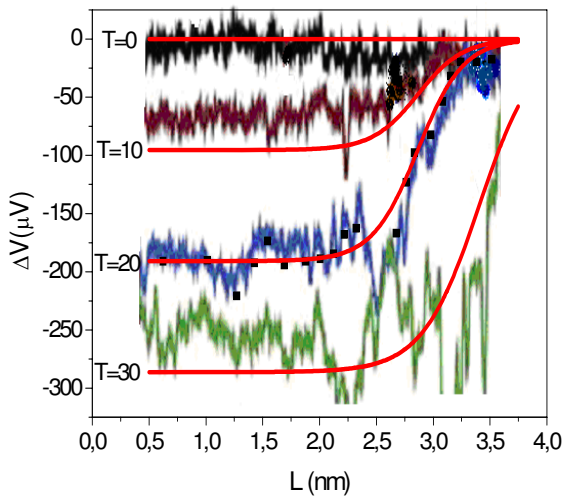}
\vskip-3mm\caption{Dependence of $\Delta V$ on the distance between
electrodes at different temperatures. Results of calculation are
compared with experiment \cite{2}.  }\label{fig:3}
\end{figure}
with an \emph{externally imposed current} $J_{sd} =\Delta V/R_0 $, where $R_{0}$ is the external
impedance of the whole measurement system. We note that if the conditions $e_0|\Delta V|\ll k_B\, T$ and
$|E_\alpha -\mu|\ll k_B\,T$ are valid, \ref{eq8} can be rewritten as
\begin{equation}
\label{eq9} (E_\alpha -\mu )\Delta T/T-e_0\Delta V=p\Delta V ,
\end{equation}
where $p=4\hbar k_B\, T/e_0R_0\, C_\alpha \,G(E_\alpha )$. In the case of an
open circuit $R_0 =\infty $, we have $p = 0$, and \ref{eq7} and
\ref{eq9} are the same. The quantities $\Gamma _{L}$ and $\Gamma _{R
}$ entering $G(E)$ are exponentially dependent on the barrier widths, i.e.\ we can
approximate them as
\begin{equation}
\label{eq10} \Gamma _{L,R} =\Gamma _0 \,\exp (-\gamma a_{L,R} ) ,
\end{equation}
where $a_{L,R}$ are the widths of left and right barriers,
respectively, and \textit{$\gamma$} depends on the barrier height. The barrier lengths can in the experimental setup be controlled by moving the STM tip \cite{2}. 
Finally, relation \ref{eq9} can be written in the form
\begin{equation}
\label{eq11} e_0\Delta V=\frac{(E_\alpha -\mu )}{1+B(d)}\,\frac{\Delta T}{T}
,
\end{equation}
where
\[
B(d)=B_0 [\exp (\gamma a_L )+\exp (\gamma a_R )],
\]
\[
B_0 =\frac{4\hbar k_B T}{e_0R_0 C_\alpha \Gamma _0 }.
\]
Note that the total distance between source and drain is $d=a_L +a_R +a_{\rm BDT} $,
where $a_{\rm BDT}$ is the diameter of the molecule BDT.

This theoretical length dependence of the thermoelectrical voltage can be
compared with experimental data \cite{2}. The result is shown
in Fig.\ \ref{fig:3}. In our calculations, we fixed the parameters by obtaining the best fit for $\Delta T=20$K, resulting
in  $a_L \approx 0.2$ nm, $a_{\rm BDT} =0.5$ nm, $\gamma =5$
nm$^{-1}$, $B_0 =2\cdot 10^{-5}$, $(E_\alpha -\mu )/T=9.54~\mu $V/K. The remaining curves for the other temperatures were then calculated using these model parameters. The general agreement between theory and experiment is very good, only at larger $\Delta T$ we observe stronger deviations. Since our model is comparatively simple and naturally contains various assumptions, such a deviation is not surprising but to be expected. It can be attributed to various sources, ranging from neglected barrier height fluctuations to influence of additional molecular orbitals or even molecular vibrations \cite{ramsak,ora,sergei}.


\section{Summary}
A theoretical description of transport through nano-structures makes a full quantum-mechanical description of the system mandatory. In contrast to bulk materials, one cannot even adopt some semi-classical approach based on e.g.\  Boltzmann equations here. Since one also needs to take into account the inherent non-equilibrium situation in many cases, solving this problem has become one of the most challenging tasks in modern condensed matter theory. A certain simplification arises when one can use
the concept of resonant tunneling. This is usually possible in weakly contacted nano-objects like molecules, and allows  to quite
accurately describe the thermoelectric phenomena in these systems.

We have presented here the calculation of thermotransport through a BEDT molecule contacted with a metallic substrate and a STM tip via generation of a break junction \cite{2}. In the limit of only weakly transparent barriers we were able to obtain an explicit formula for the voltage drop across the molecule as function of temperature difference between substrate and tip. We found that the experimental data are well described by a resonant tunneling process involving the HOMO of the molecule. 

Modelling the dependence of tunneling rates between the molecule and substrate/tip with a simple exponential ansatz, we were furthermore able to give a closed expression for the dependence of the thermoelectric effect on the distance between tip and substrate. The comparison to experiment could be done by extracting the relevant model parameters from one set of data for a fixed $\Delta T=20$K only, reproducing the remaining curves with good accuracy. Thus, our formula describes very accurately the transport through such a nano-object, provided the coupling to the leads is sufficiently weak.

There are, of course, several severe simplifications in the model. The possibly most relevant is the neglect of interaction effects on the molecule, as well as molecular vibrations and also vibrations of the whole molecule between the contacts. These additional features can be taken into account \cite{ramsak,sergei}, and under very simplifying assumptions also solved analytically \cite{sergei}.
We therefore believe that such results, may they seem  simple or straightforward, are nevertheless very important steps to enhance our knowledge about transport through nano-structures and can actually also serve as benchmarks to test more elaborate theoretial tools.


\begin{acknowledgements}
We acknowledge financial support by the Deutsche Forschungsgemeinschaft
through  project PR 298/12 (T.P. , S.K. and V.K.) .
S.K.\ furthermore thanks the Institute for Theoretical Physics of the University of G\"ottingen and T.P. the Bogolyubov Institute for Theoretical Physics of the NASU  in Kiev for their support and hospitality during the respective visits.S.K. is grateful  for a financial  support of the SASIIU of the project Ukraine - Republic of Korea.
\end{acknowledgements}


\end{document}